\documentclass[runningheads]{llncs}

\usepackage{graphicx}
\usepackage{xpatch}
\usepackage[normalem]{ulem}

\usepackage{amsmath,amssymb,amsfonts}
\usepackage{graphicx}
\usepackage{xcolor}
\usepackage{epsfig}
\usepackage{xspace}
\usepackage{graphicx}
\usepackage{subfig}
\usepackage{balance}
\usepackage{caption}
\usepackage{multirow}
\usepackage{url}
\usepackage{enumitem}
\usepackage{threeparttable}

\usepackage[ruled,noend,linesnumbered]{algorithm2e}
\let\oldnl\nl
\newcommand{\nonl}{\renewcommand{\nl}{\let\nl\oldnl}}

\usepackage{todonotes}
\makeatletter
\DeclareRobustCommand*\cal{\@fontswitch\relax\mathcal}
\makeatother

\xpatchcmd{\@thm}{\fontseries\mddefault\upshape}{}{}{}

\newcommand{\kw}[1]{{\ensuremath {\mathsf{#1}}}\xspace}

\newcommand{\stitle}[1]{\noindent{\bf #1}}

\newcommand{\newtext}[1]{{\noindent \color{black}{#1}}}

\newcommand{\lit}[1]{{\sl #1}}

\newcommand{\owl}{\kw{OWL}}

\newcommand{\gfds}{\kw{GFDs}}

\newcommand{\gfd}{\kw{GFD}}
\newcommand{\gkey}{\kw{GKey}}

\newcommand{\gkeys}{\kw{GKeys}}

\newcommand{\isounit}{\kw{IsoUnit}}

\newcommand{\gkminer}{\kw{GKMiner}}
\newcommand{\gkminerNoOpt}{\kw{GKMiner}-\kw{NoOpt}}
\newcommand{\vickey}{\kw{VICKEY}}
\newcommand{\sakey}{\kw{SAKey}}

\newcommand{\gkdiscovery}{\kw{Discovery}}

\newcommand{\dbpedia}{\kw{DBpedia}}
\newcommand{\dbpyago}{\kw{DBpediaYago}\xspace}
\newcommand{\imdb}{\kw{IMDB}}
\newcommand{\yago}{\kw{Yago}}

\newcommand{\supp}{\kw{sup}}
\newcommand{\type}{\kw{type}}
\newcommand{\cnt}{\kw{count}}
\newcommand{\prune}{\kw{prune}}
\newcommand{\unique}{\kw{unique}}
\newcommand{\bounded}{$k$-\kw{bounded}}
\newcommand{\lattice}[1]{$\mathcal{L}(#1)$}
\newcommand{\depgraph}{$\mathcal{D}$}
\newcommand{\false}{\kw{false}}
\newcommand{\true}{\kw{true}}
\newcommand{\id}{\kw{id}}

\newcommand{\eat}[1]{}

\newcommand{\thesis}[1]{}

\makeatletter
\DeclareRobustCommand*\cal{\@fontswitch\relax\mathcal}
\makeatother

\newcommand{\bi}{\begin{itemize}}
	\newcommand{\ei}{\end{itemize}}
	{\end{itemize}} 

\newcommand{\be}{\begin{enumerate}}
	\newcommand{\ee}{\end{enumerate}}
\newcommand{\beqn}{\begin{eqnarray*}}
	\newcommand{\eeqn}{\end{eqnarray*}}

\newcommand{\ie}{\emph{i.e.,}\xspace}
\newcommand{\eg}{\emph{e.g.,}\xspace}

\newcommand{\eop}{\hspace*{\fill}\mbox{$\Box$}}

\newcommand{\nthesection}{\arabic{section}}

 \newcounter{axiom}
 \renewcommand{\theaxiom}{\arabic{axiom}}

\newcounter{cor}
\renewcommand{\thecor}{\arabic{cor}}

\newcounter{prop}
\renewcommand{\theprop}{\arabic{theorem}}
 
\newcounter{alg}[section]
\renewcommand{\thealg}{\nthesection.\arabic{alg}}

\newcounter{arule}
\renewcommand{\thearule}{\arabic{arule}}

\newenvironment{proofS}{
	{\noindent\bf Proof sketch:\ }}{\eop
	}

\definecolor{gray}{rgb}{0.5,0.5,0.5}

\newcounter{ccc}

\newenvironment{ab}
{\mathactivatecomma
	\mathcode`\,=\string"8000
	\ignorespaces}
{\ignorespacesafterend}

\newcommand{\mathactivatecomma}{%
	\begingroup\lccode`~=`\,
	\lowercase{\endgroup\edef~}{\mathchar\the\mathcode`\,\penalty0 }}

\newcommand{\bab}{\begin{ab}}
	\newcommand{\eab}{\end{ab}\xspace}

\renewcommand{\prec}{\kw{precision}}
\newcommand{\rec}{\kw{recall}}

\newcommand{\fscore}{\kw{F_1}}

\begin{document}

\title{Discovery of Keys for Graphs [Extended Version]
}
%
%
\author{Morteza Alipourlangouri, Fei Chiang
}
\authorrunning{M. Alipourlangouri and F. Chiang}
%

\institute{McMaster University, 1280 Main St W, Hamilton, ON L8S 4L8 \\ \email{\{alipoum, fchiang\}@mcmaster.ca} 
}

%
\maketitle              

\begin{abstract}
\vspace{-2ex}
Keys for graphs uses the topology and value constraints needed to uniquely identify entities in a graph database. They have been studied to support object identification, knowledge fusion, data deduplication, and social network reconciliation. In this paper, we present our algorithm to mine keys over graphs. Our algorithm discovers keys in a graph via frequent subgraph expansion. We present two properties that define a meaningful key, including \emph{minimality} and \emph{support}.  Lastly, using  real-world graphs, we experimentally verify the efficiency of our algorithm on real world graphs.

\end{abstract}

\section{Introduction}
\label{sec:intro}

Keys are a fundamental integrity constraint defining the set of properties to uniquely identify an entity. Keys serve an important role in relational, XML and graph databases to maintain data quality standards to minimize redundancy and \newtext{to prevent} incorrect insertions and updates.  In addition, keys are helpful for deduplication (also referred to as entity resolution) and have been widely studied for entity identification~\cite{dong2014data,akhtar2010constraints,atencia2014defining}.  While keys are often defined by a domain analyst according to application and domain requirements, manual specification of keys is expensive and laborious for large-scale datasets.  
Existing techniques have explored mining for keys in relational data (as part of functional dependency discovery)~\cite{huhtala1999tane}, and in XML data~\cite{buneman2002keys}.  

The expansion of graph databases has lead to the study of integrity constraints over graphs, including  functional dependencies~\cite{fan2017dependencies,alipourlangouri2021temporal}, keys~\cite{fan2015keys} and their ontological invariant~\cite{ma2019ontology}. The theoretical foundation of these constraints have been \newtext{studied} and there has been a wide application of key constraints for deduplication, citation of digital objects, data validation and knowledge base expansion~\cite{dong2014data,hellings2014implication}. Graphs such as knowledge bases and citation graphs require keys to uniquely identify objects to ensure reliable and accurate deduplication and query answering. There is a need to automatically discover keys from such graphs as manual specification of keys is expensive and labor intensive.  Although recent work has proposed techniques to find keys over RDF data \cite{atencia2014defining}, these techniques are not applicable for graphs as they do not support: (i) topological constraints; and (ii) recursive keys (a distinct feature in graph keys).  
Consider the following example on how keys help us to identify entities in a graph.

\vspace{-2ex}
\begin{figure}[ht!]
\centering
\includegraphics[width=4.8in,keepaspectratio]{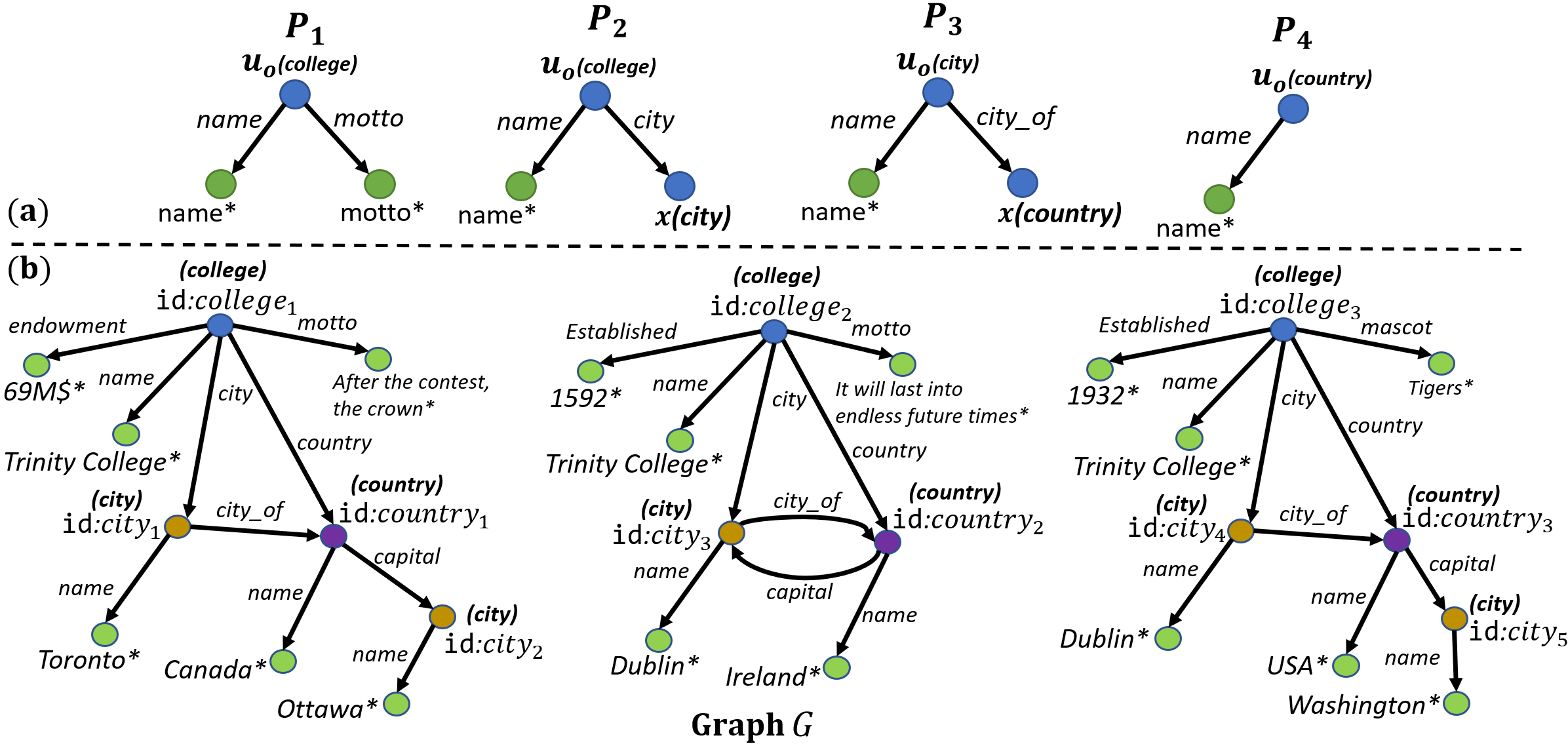}
\caption{Sample graph from \dbpedia.}
\label{fig:mainGraph}
\end{figure}
\vspace{-2ex}

\begin{example} Consider a knowledge graph consisting of triples (subject, predicate, object) where subject and object are nodes, and predicate is an edge connecting subject to
object. Figure \ref{fig:mainGraph} shows a sample of such graph from the \dbpedia dataset~ \cite{lehmann2015dbpedia} of three colleges, five cities, and three countries along with the attributes of each entity. Consider graph keys with patterns $P_1$ and $P_2$ in Figure~\ref{fig:mainGraph}. $P_1$ states that \emph{if two colleges share the same name and motto, then they refer to the same college}. $P_2$ states that \emph{if two colleges share the same name and city, then they refer to the same college.}, Similarly, a city can be identified by its name and country as it is shown in $P_3$. Moreover, $P_4$ states that \emph{a country can be identified by its name}. Note that $P_2$ is dependant to $P_3$ and $P_3$ is dependant \newtext{on} $P_4$, which reflects the recursiveness of graph keys~\cite{fan2015keys}. \thesis{However, one can confirm that not all the attributes can form a graph key for an entity. For example, \emph{name} cannot uniquely identify \emph{college} (resp. \emph{city}) as we have three colleges (resp. two cities) with the same name.}
\end{example}

The example highlights that many keys are possible to identify entities, and this depends on the data and its semantics. The domain semantics influence the quality of a key.  For example, $P_1$ uses the \emph{name} and \emph{motto} to uniquely identify the college.  However, not all colleges have motto, and this lead us to null values for some colleges, thereby leading to poor support and representation across all colleges. This highlights the need to define meaningful properties for a key and an efficient automatic discovery of such keys over graphs.

\noindent \textbf{Contributions.} (1) We define new properties for graph keys (\emph{support} and \emph{minimality}), and formalize the graph key discovery problem.  (2) We introduce \gkminer, an algorithm that mines all recursive graph keys by using novel auxiliary structures and optimizations to prune unlikely key candidates.  (3) Lastly, we evaluate \gkminer over two real data graphs, and show its scalability and efficiency over three baselines.

The rest of the papers is structured as follows. We discuss related works in Section~\ref{sec:relatedWorks}, and preliminaries in Section~\ref{sec:preliminaries}. In Section~\ref{sec:GKeyDiscovery} we provide key properties and then the discovery algorithm. We present our experimental evaluations in Section~\ref{sec:exp}. Finally, we have conclusion and future works in Section~\ref{sec:future}.

\section{Related Works}
\label{sec:relatedWorks}

\stitle{Keys and Dependencies.}
Keys are defined to uniquely identify entities in a database. For relational data, keys are defined as a set of attributes over a schema~\cite{abiteboul1995foundations}, or by using unique column combinations~\cite{birnick2020hitting,wei2019discovery} to uniquely identify the tuples. For XML data, keys are defined based on path expressions in the absence of schema~\cite{buneman2002keys}. Traditional keys are also defined over RDFs~\cite{atencia2012keys,pernelle2013automatic,soru2015rocker} in the form of a combination of object properties and data properties defined over \owl \ ontology.
Recent works have studied functional dependencies for graphs (\gfds) that define value constraints on entities that satisfy a topology constraint~\cite{fan2019dependencies,fan2016functional}. Keys for graphs (\gkeys) aim to uniquely identify entities represented by vertices in a graph, using the combination of
recursive topological constraints and value equality constraints. \gkeys \ are a special case of \gfds~\cite{fan2015keys}. The recursiveness of \gkeys \ makes it more complex compare to the relational and RDF based counterparts. Graph matching keys, referred to as \kw{GMKs}, are extension of graph keys using similarity predicates on values, and supporting approximation entity matching~\cite{deng2020keys}. 

\kw{PG}-\kw{Keys}~\cite{angles2021pg} proposes a modular and flexible model to formalise keys for property graph. Their keys are defined to be used for a property graph query language that is currently underway through
the ISO Graph Query Language (GQL) project. \kw{PG}-\kw{Keys} focuses to define keys that are applicable to nodes, edges, and properties in a property graph. However, they do not consider topology constraint to define a key and consider schema to define keys, while \gkeys are focused to uniquely identify entities (\ie nodes) in the absence of schema. For the property graphs, a uniqueness constraint is a set of attributes whose values uniquely identify an entity in the collection. Neo4j keys~\cite{link2020neo4j} are based on uniqueness constraints and require the existence of such constraints for all vertices the graph. A new principled class of constraints called embedded uniqueness constraints have been proposed that separates uniqueness from existence dimensions and are used in the property graphs to uniquely identify entities~\cite{skavantzos2021uniqueness}. However, \gkeys are different than these constraints by supporting topological constraint through graph pattern. 

\stitle{Dependency Discovery and Pattern Mining}. Key mining approaches have been studied for relational databases as data-driven~\cite{heise2013scalable} and schema-based~\cite{sismanis2006gordian} techniques. TANE~\cite{huhtala1999tane} proposed a level-wise schema-based approach to mine keys in relational data (as part of functional dependency) and it has been extended for RDFs~\cite{atencia2012keys}.
KD2R~\cite{pernelle2013automatic} extends the relational data-driven approach
of~\cite{sismanis2006gordian} by exploiting axioms (such as the subsumption relation) and
considers multi-valued properties. SAKey~\cite{symeonidou2014sakey} extends K2DR by introducing additional pruning techniques to discover approximate keys with exceptions. VICKEY~\cite{symeonidou2017vickey} has extended SAKey to mine conditional keys over RDFs. To avoid scanning the entire dataset, all three techniques (\ie K2DR, SAKey, and VICKEY) first discover the maximal non-keys and then derive the keys from this set. Non-keys are the set of attributes that are not keys and maximal non-keys are super-sets of all other non-keys. Instead of exploring the whole set of combinations of properties, the idea behind these techniques is to find those combinations that are not keys and then derive the keys from that set. Fan et. al, have developed a parallel algorithm to discover \gfds \ in graphs~\cite{fan2020discovering}. Although \gkeys are a special case of \gfds, their technique is not able to mine \gkeys. In order to model \gkeys as \gfds, we need to have a graph pattern consist of two connected components to define equality over pairs of matches. However, \gfd \ discovery algorithm~\cite{fan2020discovering} only mines \gfds \ with a single connected component pattern, which makes it impossible to mine \gkeys \ using their technique. To the best of our knowledge, there is only one technique to discover keys for graphs~\cite{alipourlangouri2018keyminer}, which is our preliminary work published at VLDB-TDLSG workshop. This work differs from~\cite{alipourlangouri2018keyminer} as we define new metrics to mine \gkeys, and propose an efficient algorithm with optimization and perform extensive experimental evaluations over real world graphs and compare with \sakey~\cite{symeonidou2014sakey}.

\vspace{-3ex}
\section{Preliminaries}
\label{sec:preliminaries}
\vspace{-2ex}

\stitle{Graphs.} A \emph{directed graph} is defined as $G=(V,E,L,F)$ with labeled nodes and edges, and attributes on its nodes. The set $V$ is a finite set of vertices and $L$ is a finite set of labels.  A set of edges is denoted as $E \in V \times L \times V$, \ie $e=(u,l,v)$ represents an edge from $u$ to $v$ with the label $l$ that is not equal to edge $(v,l,u)$. Each node $v \in V$ may have a label $l \in L$ referred as  $v.\type$ and a numeric \id, denoted by $v.\id$.
For a node $v$, $F(v)$ is a tuple to specify the set of attributes as $(A_1 = a_1, . . . , A_n = a_n)$ of $v$. More specifically, $A_i$ with a constant $a_i$ determines the attribute $A_i$ of $v$ written as $v.A_i = a_i$. Attributes can carry the properties of a node such as \lit{name}, \lit{age}, etc., as found in social networks and knowledge graphs. We represent each attribute as a separate \newtext{node} with no $\type$ and $\id$ in our graph \ie for each attribute $(A_i = a_i) \in F(v)$, there exists a node $v_i$ with the value of $a_i$ and there exists a corresponding edge $(v,A_i,v_i) \in G.E$.

\thesis{\begin{example}
We return to Figure~\ref{fig:mainGraph}, where we have three colleges with unique ids $\{college_1, college_2, college_3\}$ and all of them have the property \lit{name = Trinity College}. However, $college_1$ and $college_2$ have \lit{motto}, while $college_3$ has \lit{mascot}. Moreover, entities are connected to each other via edges, \eg $city_1$ with the \lit{name = Toronto}  \newtext{has an edge \emph{city\_of} to} $country_1$ named \lit{Canada}.
\end{example}
}

\stitle{Graph pattern.} A \textit{graph pattern} is defined as a connected, directed graph $P(u_o)=(V_P, E_P, L_P)$ where (1) $V_P$ is a finite set of pattern nodes; (2) $E_P$ is a finite set of pattern edges; (3) $L_P$ is a function which assigns a specific label $L_P(v)$ (resp. $L_P(e)$) to each vertex $v\in V_P$ (resp. each edge $e \in E_P$).
The pattern nodes $V_P$ may be one of three types: (1) a {\em center} node $u_o\in V_P$, representing the main entity to be identified; 
(2) a set of {\em variable nodes} $V_x\subseteq V_P$; and (3) a set of {\em constant nodes} $V_c$ = $V_P\setminus (\{u_o\}\cup V_x)$. A variable node is being mapped to an \emph{entity} and it carries the label as a type along with an \id, while a constant node only contains a value without any \id to map to a value.

\thesis{
\begin{example}
Consider the two entity patterns $P_1$ and $P_2$ in Figure~\ref{fig:mainGraph}(a), characterizing entities of type \lit{college}. $P_1$ contains constant nodes \lit{name} and \lit{motto}, while $P_2$ has a constant node \lit{name} and a variable node \lit{city}. $P_3$ is a pattern for the entities of type \lit{city} with constant node \lit{name} and variable node \lit{country}, while \lit{country} has pattern $P_4$ with a constant node \lit{name}.
\end{example}
}

\stitle{Graph pattern matching.}
Given two labels $\iota$ and $\iota'$ from $L_P$, we say $\iota$ matches $\iota'$, denoted as $\iota \asymp \iota'$
if either (1) $\iota = \iota'$; (2) $\iota=`\_$', \ie wildcard matches any label. 
Given a graph $G$ and a pattern $P(u_o)$, \newtext{a match $h$ is a subgraph $G'=(V',E',L',F_A')$, which is isomorphic to $P$, \ie there exists a bijective function $h$ from $V_P$ to $V'$ such that} (i) for each node $v \in V_P$, $L_P(v) \asymp L'(h(v))$; and (ii) for each edge $e(u,u') \in E_P$, there exists an edge $e'(h(u),h(u')) \in G'$ such that $L_P(e)=L'(e')$.

\begin{example} \label{exm:match}
Given pattern $P_1$ of Figure~\ref{fig:mainGraph}(a), we can find matches $h_1$ and $h_2$ in graph $G$ of Figure~\ref{fig:mainGraph}(b), such that $h_1(college)=college_1$ and $h_2(college)=college_2$. $college_3$ is not a match of $P_1$ as there is no match for the node \lit{mascot}. However, there exist three matches $h_1$, $h_2$ and $h_3$ for pattern $P_2$ in $G$ for $college_1$, $college_2$ and $college_2$ respectively. \newtext{Similarly, we have three cities $city_1$, $city_3$, $city_4$ matched with the pattern $P_3(city)$ in $G$ and all three countries $country_1$, $country_2$ and $country_3$ are matched with pattern $P_4(country)$.}
\end{example}

\stitle{Graph keys (\gkeys).}
A key for a graph is defined using a pattern $P(u_o)$ for a designated entity $u_o$ \cite{fan2015keys}. Given two matches $h_1$ and $h_2$ of $P(u_0)$ in graph $G$, $(h_1,h_2)$ satisfies $P(u_0)$ denoted as $(h_1,h_2)\models P(u_0)$, if (a)$\{\forall v \in V_x, h_1(v).\id = h_2(v).\id\}$; (b) $\{\forall v \in V_c,  L(h_1(v)) \asymp L(h_2(v))\}$; and (c)$\{\forall e \in E_P, L(h_1(e))=L(h_2(e))\}$; then $h_1(u_o).\id = h_2(u_o).\id$.  This means the two matches refer to the same entity in $G$. We say a graph $G$ satisfies a key $P(u_o)$, denoted as $G \models P(u_o)$, if for every pair of matches \newtext{$(h_1,h_2) \in G$}, we have $(h_1,h_2)\models P(u_0)$.
Moreover, \newtext{a} key $P(u_o)$ is considered as a \emph{recursive} key if it contains \newtext{at least one variable $v \neq u_o$, otherwise, $P(u_o)$ is called a \emph{value-based} key~\cite{fan2015keys}}.

\begin{example}
Going back to Figure~\ref{fig:mainGraph}, a \gkey $P_1(college)$ can uniquely identify $college_1$ and $college_2$ as they have different \lit{motto}, despite the same \lit{name}. $P_2(college)$ is a recursive \gkey that can identify all three colleges. It is recursively dependant to \lit{city} of $P_3(city)$, while \lit{city} is recursive defined via \lit{country} of the \gkey $P_4(country)$. Although $city_3$ and $city_4$ have the same name \lit{Dublin}, but they belong to different countries \lit{USA} and \lit{Ireland}\newtext{,} respectively. Two level of recursions help $P_2(college)$ to uniquely identify all three colleges in $G$.
\end{example}

\section{Discovery of \gkeys}
\label{sec:GKeyDiscovery}

In this section, we discuss the discovery problem for \gkeys. The discovery problem is to find a set of \gkeys  for a given type $u_o$ in an input graph $G$. Graph keys are able to impose topological constraint along with attribute value bindings that are needed to identify entities. Existing works miss the topology and only discover keys as a set of attribute value that work over RDF data. While we mine keys by considering both topology and attribute values in the form of a graph pattern~\cite{fan2015keys} However, it is not desirable to mine all \gkeys  for $u_o$ as a large amount of them are redundant and not meaningful. Mining meaningful keys in graphs relies on defining key properties independent of the application domain. We propose two key properties: \emph{minimality} and \emph{support}, and a key discovery algorithm over graphs. 

\subsection{Key Properties} \label{sec:GKeyProperties}
We now present our approach to mine all minimal \gkeys  $\Sigma$ in $G$ for a given entity type $u_o$ such that $G \models \Sigma$. \emph{Minimality} avoids mining redundant \gkeys  and reduces the discovery time. \emph{Support} mines keys that satisfy minimum number of instances in $G$. We first introduce notion of \gkey  embedding.

\stitle{\gkey  embedding}. We say a \gkey  $P(u_o) = (V_P, E_P, L_P)$ is {\em embeddable} in another \gkey  $P'(u_o) = (V_P', E_P', L_P')$, if there exists a subgraph isomorphic mapping $f$ from $V_P$ to a subset of nodes in $V_P'$ that preserves node labels/values of $V_P$, and all the edges that are induced by $V_P$ with the corresponding edge labels. 

\stitle{Minimality.} A \gkey $P(u_o)$ is minimal if there exists no \gkey  $P'(u_o)$ such that $P'(u_o)$ is embeddable in $P(u_o)$. A set $\Sigma$ of \gkeys  with $G \models \Sigma$ is minimal, if it does not contain any redundant \gkeys. A redundant \gkey  $P(u_o)$ exists in $\Sigma$, if removing $P(u_o)$ from $\Sigma$ results
in a $\Sigma'$ that is logically equivalent to $\Sigma$, \ie $\Sigma'$ uniquely identifies the same entities as $\Sigma$ in $G$.

\stitle{Support.} For a candidate \gkey  $P(u_o)$, we define support to represent the number of entities in the graph $G$ that are uniquely identified by $P(u_o)$ over the total number of entities of type $u_o$. We define $|P(u_o)|$ as the total number of entities that are uniquely identified by $P(u_o)$. for a \gkey  $|P(u_o)|$ such that $G \models P(u_o)$, we 
define $\supp(P(u_o)) = \dfrac{|P(u_o)|}{N}$, where $N$ be the total number of instances with the type $u_o$ in graph $G$.

\thesis
{\begin{equation} \label{eq:coverage}
    \kw{sup}(P(u_o)) = \dfrac{|P(u_o)|}{N}
\end{equation}
Let $N$ be the total number of instances with the type $u_o$ in graph $G$.
}

\stitle{\emph{k-bounded} \gkeys.} For a given user defined natural number $k$, a \gkey  $P(u_o)$ is \emph{k-bounded} if $\kw{size}(P(u_o))\leq k$, where $\kw{size}$ is defined as: 

\begin{equation} \label{eq:sizeOfKey}
    \kw{size}(P(u_o)) = |E_p| + \kw{size}(P(v_P))  \forall v_p \in V_P
\end{equation}

This equation counts the number of edges in the pattern $P(u_o)$ and in the pattern of all variable nodes \ie recursive \gkey. To validate a recursive \gkey, one must validate the matches of the recursive patterns\cite{fan2015keys}. A set $\Sigma$ of \gkeys  is \emph{k-bounded}, if each $P(u_o) \in \Sigma$ is \emph{k-bounded}.

\stitle{Problem statement.} Given a graph $G$, a node type $u_o$, a support threshold $\delta$, and a natural number $k$, mine all minimal \emph{k-bounded} \gkeys  $\Sigma$ of the node type $u_o$, such that for each \gkey  $P(u_o) \in \Sigma$, $P(u_o)$ has the minimum support $\delta$ in $G$.

\subsection{Algorithm} \label{sec:GKeyAlgorithm}

For a given entity type $u_o$, the naive algorithm mines all frequent graph patterns centered by $u_o$ and explores all combinations of variable and constant nodes in each pattern to verify whether they form a \gkey. The naive approach leads us to explore a large search space, which is shown to be infeasible in real world graphs~\cite{fan2018discovering}.
We introduce \gkminer, an efficient algorithm to mine all minimal \gkeys  in a graph. Our algorithm takes as input a graph $G$, an entity type $u_o$, a natural number $k$ and a support threshold $\delta$ to discover \gkeys. It proceeds in three steps: (a) Create a summary graph $\mathcal{S}$ to explore the structure of $G$. This will help us to prune nodes that cannot form a \gkey  based on the given threshold $\delta$. (b) Create a lattice $\mathcal{L}$ of candidate \gkeys  from $\mathcal{S}$ that prunes further candidate \gkeys. (c) Mine minimal \emph{k-bounded} \gkeys  from $\mathcal{L}$ in a level-wise search. Using lattice to model candidate \gkeys, and with early pruning techniques performed on the lattice, \gkminer is able to avoid the large search space of the naive approach. Our experiments show that our algorithm runs up to 6 times faster than \sakey, despite mining topological constraints of \gkeys compared to the value constraint based keys mined by \sakey.

\stitle{Summary Graph.} As the first step of mining \gkeys, we traverse $G$ to create a summary graph $\mathcal{S}$ that reflects the structure of $G$.  $\mathcal{S}$ provides an abstract graph of $G$, where: (1) nodes represent the entity types that exist in $G$, and (2) an edge between two nodes in $\mathcal{S}$ shows that there exists at least one edge between two entities with the corresponding types in $G$. $\mathcal{S}$ helps us to model the relationship between entity types in a smaller graph. $\mathcal{S}$ will be used to define graph patterns for candidate \gkeys. $\mathcal{S}$ is built in $O(V+E)$ time and is an auxiliary data structure $\mathcal{S}(V_S,E_S)$, where $V_S$ (resp. $E_S$) is a set of nodes (resp. edges) and have the following properties:
\begin{enumerate}
    \item for each node type $t \in L$ in the graph $G$, there exists a node $v_t$ in $V_S$.
    \item For each node $v_t \in V_S$, $v_t.\cnt$ is the number of nodes in $G$ of type $t$.
    \item For each edge $e(u_1,l_e,u_2) \in G$:
    \begin{enumerate}
        \item if $u_1$ is of type $t_1$ and $u_2$ does not carry a type, \ie a constant node, then create an attribute $A_e$ with the name $l_e$ and without any value (\eg set value as $*$) and add to $v_{t_1}$ in $V_S$.  Increase the $A_e.\cnt$ by one (initial value is 0).
        \item if $u_1$ is of type $t_1$ and $u_2$ is of type $t_2$, then add an edge $e(v_{t_1},l_e,v_{t_2})$ to $E_S$ and increase the $e.\cnt$ by one (initial value is 0).
    \end{enumerate}
\end{enumerate}

\begin{figure}[ht!]
\centering
\includegraphics[width=4.8in,keepaspectratio]{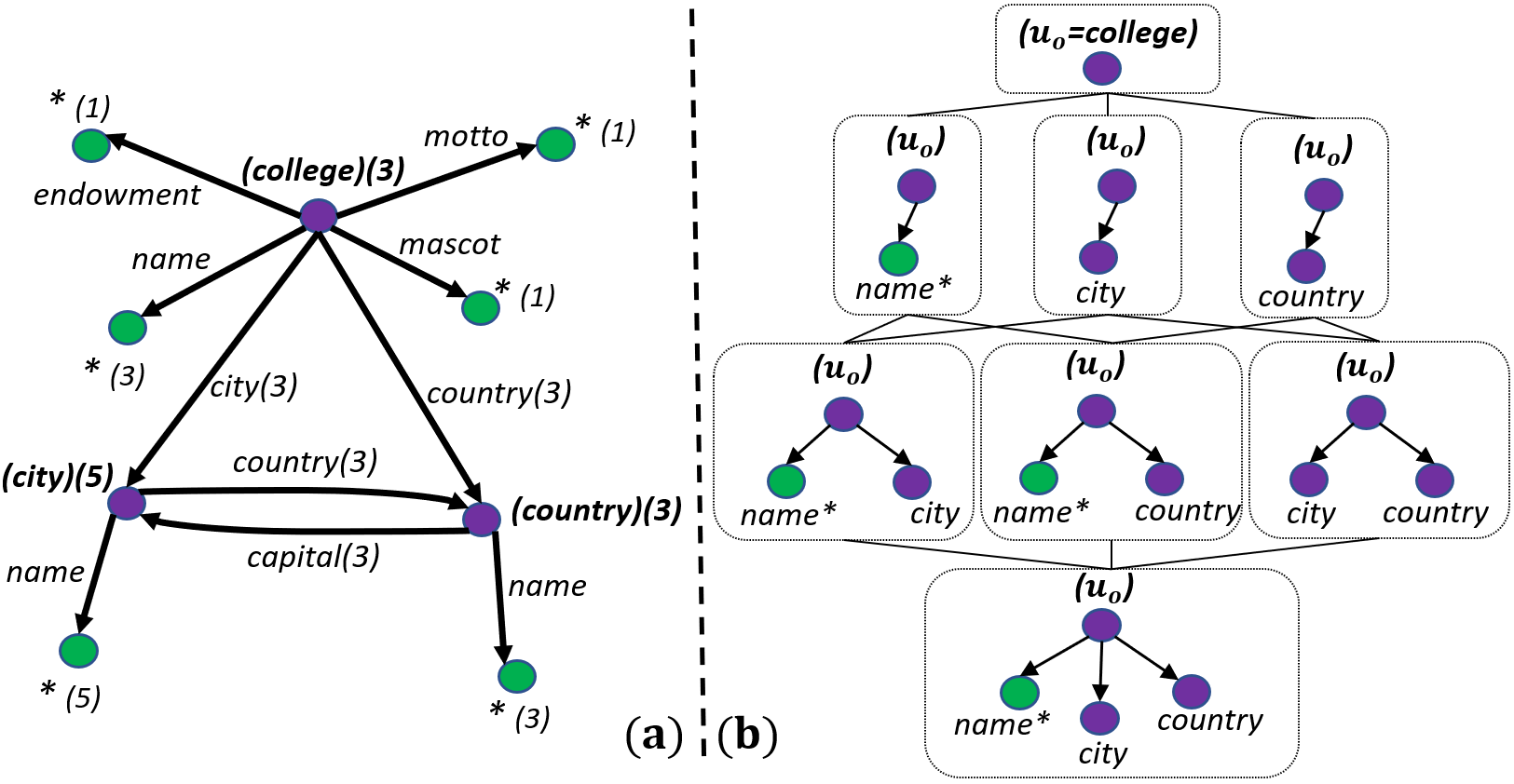}
\caption{(a) Summary graph $\mathcal{S}$ of graph $G$ of Figure~\ref{fig:mainGraph} (b) Lattice for the type $college$ based on $\supp=75\%$}
\label{fig:summaryGraph}
\end{figure}

\begin{example}
Figure~\ref{fig:summaryGraph}(a) shows the summary graph generated for the graph $G$ of Figure~\ref{fig:mainGraph} where we have three entities of type \lit{college},  five entities of type \lit{city} and three of type \lit{country}. Out of three colleges, one of them has the attribute \lit{endowment}, one has \lit{mascot}, two have \lit{motto} and all three have attribute \lit{name}.
\end{example}

After computing the summary graph, we prune the summary graph to find a set of attributes and variable nodes that meet the support threshold $\delta$, if they are added to a candidate \gkey. For a given summary graph $\mathcal{S}(V_S,E_S)$, and a given center node $v_{u_o}$, we first prune the attributes of $v_{u_o}$ based on $\delta$. Following the support definition of Section~\ref{sec:GKeyProperties}, we compute the support of an attribute $A$ of $v_{u_o}$ in $V_S$ as following:

\begin{equation} \label{eq:attributeSupport}
 \supp(A) = \dfrac{A.\cnt}{v_{u_o}.\cnt}   
\end{equation}

This equation computes the support of an attribute $A$, if we add $A$ as a singleton attribute in a candidate \gkey. If we have $\supp(A)<\delta$, then adding $A$ to any candidate \gkey  $P(u_o)$,  makes $\supp(P(u_o))<\delta$, hence $P(u_o)$ will not be a valid \gkey. 
Therefor, we select a set of candidate attributes $\mathcal{A} = \{A_1,\dots, A_n\}$ of $v_{u_o}$ in $V_S$ such that for each $A_i$, $\supp(A_i) \geq \delta$. 

Similar to the Equation~\ref{eq:attributeSupport}, we compute the support of the variable nodes, that are immediate neighbors of $v_{u_o}$. If $v_{u_o}$ is connected to a node $v$ with an edge $e$, then the support of $v$ is computed as:

\begin{equation}\label{eq:variableNodeCoverage}
    \supp(v) = \dfrac{e.\cnt}{v_{u_o}.\cnt}     
\end{equation}

Following the same reasoning of Equation~\ref{eq:attributeSupport}, adding a variable node $v$ with $\supp(v)< \delta$ to a \gkey  $P(u_o)$, makes $\supp(P(u_o))< \delta$. Hence,  we define a set of variable nodes $\mathcal{V} = \{v_1,\dots, v_n\}$, where $v_{u_o}$ is connected to each $v_i$ and $\supp(v_i)\geq \delta$.

\stitle{Lattice $\mathcal{L}$.}
For the entity type $u_o$, we create a lattice \lattice{u_o} of candidate patterns based on the set $\mathcal{A}$ and $\mathcal{V}$ that are extracted from the summary graph $\mathcal{S}(V_S,E_S)$. \lattice{u_o} is rooted at node $u_o$ and expands level-wise based on the attributes in $\mathcal{A}$ and immediate variable nodes connected to $v_{u_o}$ in $\mathcal{V}$. We create the lattice \lattice{u_o} as following:

\begin{enumerate}
    \item Create a lattice \lattice{u_o} rooted at node $x$ of type $u_o$ (level 0).
    \item or the first level, we create a candidate \gkey  for the attributes and variable nodes in $\mathcal{A}$ and $\mathcal{V}$. For each attribute $A_i \in \mathcal{A}$, we create a candidate \gkey  by connecting $u_o$ to $A_i$ with an edge labeled by the name of $A_i$ and add the candidate to \lattice{u_o}. For each variable node $v_i \in \mathcal{V}$, we connect $u_o$ to $v_i$ with the corresponding edge label from $\mathcal{S}$ and add as a candidate \gkey  to \lattice{u_o}.
    \item At level $l$, we create a graph pattern for each $l$-combinations of the attributes and nodes in $\mathcal{A}$ and $\mathcal{V}$ respectively. Similarly, we connect $u_o$ to each of the nodes with a direct edge and add the pattern to $\mathcal{L}$. A candidate pattern $P(u_o)$ of level $l-1$ is connected to a pattern $P'(u_o)$ of level $l$ with a direct edge, if $P(u_o)$ is embedded in $P'(u_o)$.
    \item Each pattern $P(u_o) \in$ \lattice{u_o} has a boolean flag $P(u_o).\prune$ set by default to $\false$. This flag helps us to mine minimal \gkeys  and prune the candidates in the lattice.
\end{enumerate}
The lattice \lattice{u_o} is created for the entity type $u_o$ to generate candidate \gkeys  that initially meet the support threshold $\delta$. However, since \lattice{u_o} might contain other recursive entity types from the set $\mathcal{V}$, we need to create a lattice \lattice{v_i} for each entity type $v_i \in \mathcal{V}$.

\begin{example}
Figure~\ref{fig:summaryGraph}(b) shows the sample lattice created for the type \emph{college} based on the summary graph of Figure~\ref{fig:summaryGraph}(a) given the support threshold $\supp=75\%$. If we calculate the \supp for the attributes of the \lit{college}, we have $\supp(name)=\dfrac{3}{3}$, $\supp(endowment)=\dfrac{1}{3}$, $\supp(motto)=\dfrac{1}{3}$, and $\supp(mascot)=\dfrac{1}{3}$. Based on the \supp=75\%, we have $\mathcal{A}=\{name\}$. Similarly, if we compute the support of variable nodes connected to \lit{college}, we have $\supp(city)=\dfrac{3}{3}$, and $\supp(country)=\dfrac{3}{3}$, leads us to have $\mathcal{V}=\{city,country\}$. Using $\mathcal{A}$ and $\mathcal{V}$, we created the lattice in Figure~\ref{fig:summaryGraph}(b), where we have three levels in the lattice with seven candidates \gkeys.
\end{example}

\stitle{\gkminer  algorithm.} 
\gkminer  is a sequential \gkey  mining algorithm that traverses a lattice in a level-wise manner to mine all \gkeys  for a given type $u_o$. We first create the summary graph $\mathcal{S}$ from the input graph $G$. Next, we create the main lattice \lattice{u_o} and traverse the lattice level by level to discover \gkeys and prune when an embeddable key is already mined.
For each candidate $P_i(u_o)$ at level $i$, we check if it forms a \gkey  via incremental matching algorithm \isounit  which enables localized subgraph isomorphism~\cite{fan2013incremental}. For each candidate $P_i(u_o)$ that has the \prune  flag equal to \false, we first check $\kw{size}(P_i(u_o))$ to ensure it is \bounded. If $\kw{size}(P_i(u_o))>k$, then we set \prune=\true  for all the descendant nodes of $P_i(u_o)$ in \lattice{u_o}. Next, we calculate $\supp(P_i(u_o))$ by computing the matches as described in Section~\ref{sec:GKeyProperties}. If $\supp(P_i(u_o)) \geq \delta$, then we report $P_i(u_o)$ as a \gkey  and prune its descendant nodes in \lattice{u_o} to ensure the minimality of \gkeys. However, if $\supp(P_i(u_o)) < \delta$, we ignore $P_i(u_o)$ and continue with the next candidate.

\stitle{Handling recursive \gkeys.} In the process of mining \gkeys  for the type $u_o$, if the candidate $P_i(u_o)$ contains a variable node of type $t$ (\ie $P_i(u_o)$ is a recursive key), we first need to evaluate and find the \gkeys  for the dependant type $t$. To this end, we create the lattice \lattice{t} and recursively call the \gkminer  for the type $t$. We maintain a data structure called \emph{dependency graph} \depgraph$(V_D,E_D)$ to detect and avoid cycles in recursive calls. Cycles lead us to fall into an infinite loop of recursive calls similar to deadlocks in process management~\cite{peterson1985operating}. Cycle happens when there exists a set of types that the \gkey  of each type is dependant to the \gkey  of another type in the cycle. Using dependency graph, we follow a cycle prevention strategy and avoid cycles in recursive calls. To avoid such cycles, whenever we call \gkminer  for the type $t$ while mining \gkeys  for type $u_o$, we add $u_o$ and $t$ to $V_D$ of \depgraph  and then add a direct edge $(u_o,t)$ to $E_D$. 

\begin{algorithm}[tb!]
\kw{keys} := $\emptyset$; /* set of keys for each type*/ \\
\kw{Initialize} $\mathcal{D}(V,E) := \emptyset$  /*empty dependency graph*/ \\
\kw{Initialize} $\mathcal{S}(V,E) := \emptyset$  /*empty summary graph*/ \\
\ForEach{node $v \in G.V$}
{
    $t = v.\type$;\\
    \If{$t \neq \kw{null}$}
    {
        \textbf{if} $u_{t} \not\in \mathcal{S}.V$ \textbf{then} add $u_{t}$ to $\mathcal{S}.V$; \\ 
        $u_{t}.\cnt ++$;   
    }
}
\ForEach{edge $(v_1,l,v_2) \in G.E$}
{
    $t=v_1.\type$;\\
    \If{$v.\type == \kw{null}$}
    {
        \textbf{if} $l \not\in F(u_t)$ \textbf{then} add $l$ to $F(u_t)$; /*add $l$ as an attribute of $u_t$*/
        $u_t.l.\cnt++$;
    }
    \Else
    {
        $t'=v_2.\type$;\\
        \textbf{if} $(u_t,l,u_{t'}) \not\in \mathcal{S}.E$ \textbf{then} add $(u_t,l,u_{t'})$ to $\mathcal{S}.E$;\\
        $(u_t,l,u_{t'}).\cnt++$;
    }
}
\gkdiscovery($G$, $u_o$, $\mathcal{S}$, $\mathcal{D}$, \kw{keys}, $k$, $\delta$, $0$);\\
\KwRet{\kw{keys}};
\caption{\gkminer($G$, $u_o$, $k$, $\delta$)}
\label{alg:gkminer}
\end{algorithm}

In general, if adding an edge $(t_i,t_j)$ leads us to have a cycle in \depgraph, we break the cycle by removing the dependency $(t_i,t_j)$. To this end, we remove $t_j$ from the nodes in \lattice{t_i}. In this case, the \gkeys  of $t_i$ won't be dependant to the \gkeys  of $t_j$.

\begin{figure}[ht!]
\centering
\includegraphics[width=4in,keepaspectratio]{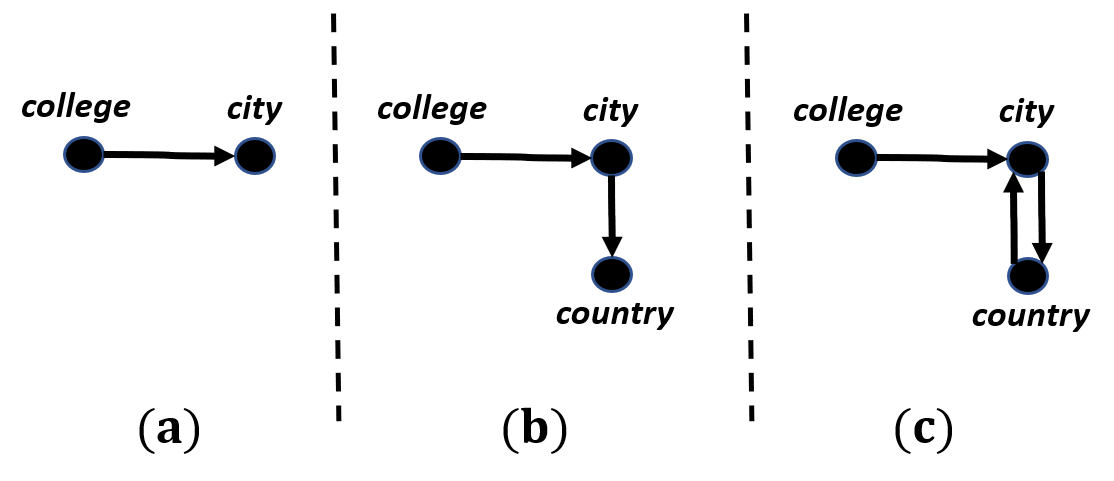}
\caption{Dependency graph $\mathcal{D}$}
\label{fig:dependencyGraph}
\end{figure}

\begin{example}
Going back to Figure~\ref{fig:summaryGraph}(b) and assuming \supp=$60\%$, when we want to check a \gkey  for the type $college$ that contains the type $city$, we find that there exists no \gkey for $city$ yet. Hence, we need to call \gkminer  for $city$ and we add an edge $(college,city)$ to the dependency graph $\mathcal{D}$ as shown in Figure~\ref{fig:dependencyGraph}(a). While mining \gkey  for $city$, we need to call \gkminer  for the type $country$ and we add an edge $(city,country)$ to $\mathcal{D}$ in Figure~\ref{fig:dependencyGraph}(b). However, while mining \gkeys  for the $country$ and as there is no \gkey  for the type $city$ yet, we cannot call \gkminer  for $city$. As shown in Figure~\ref{fig:dependencyGraph}(c), the edge $country,city$ makes a cycle in $\mathcal{D}$. Hence, we need to remove $city$ from all the candidate for the type $country$ and continue the mining to avoid cycle in $\mathcal{D}$.
\end{example}

Algorithm~\ref{alg:gkminer} provides the pseudo code of the \gkminer. After initialization (line 1-3), we create the summary graph $\mathcal{S}$ by iterating over the nodes and edges of $G$. For each node $v \in G.V$, we add a node of the corresponding type of $v$ to $\mathcal{S}$ and maintain the $\cnt$ of the nodes (lines 4-9). Next, we iterate over the edges of $G$ and add/maintain the edges and their \cnt  in $\mathcal{S}$ based on the type of the two end points of each edge in $G$ (lines 10-20). After creating the summary graph $\mathcal{S}$, we call the \gkdiscovery  algorithm and pass $\mathcal{S}$ and $u_o$ along with other inputs to find \gkeys. 

\begin{algorithm}[tb]
$\mathcal{L}(u_o) := \kw{createLattice}(\mathcal{S}, u_o, \delta)$; /*create lattice for the given type $u_o$*/ \\
\ForEach{pattern $P(u_o) \in \mathcal{L}(u_o)$}
{
    \If{$P(u_o).\prune==\false$}
    {
        \If{$|E_{P(u_o)}| + \kw{size}>k$}
        {
            $P(u_o).\prune=\true$; \textbf{continue};
        }
        \If{$P(u_o)$ contains type $t$ and $\kw{keys}[t]==\kw{null}$}
        {
            add $(u_o,t)$ to $\mathcal{D}$;\\
            \If{$\mathcal{D}$ has cycle}
            {
                remove $(u_o,t)$ from $\mathcal{D}$; remove $t$ from $P(u_o)$ and $\mathcal{L}(u_o)$;
            }
            \Else
            {
                \gkdiscovery($G$, $t$, $\mathcal{S}$, $\mathcal{D}$, \kw{keys}, $k$, $\delta$, $|E_{P(u_o)}| + \kw{size}$);\\
                \If{$\kw{keys}[t]==\kw{null}$}
                {
                    $P(u_o).\prune=\true$; \textbf{continue};
                }
            }
        }
        $\mathcal{M} := \isounit(G, P(u_o))$;\\
        $\Pi_{P(u_o)} :=\kw{computeClasses}(\mathcal{M}, P(u_o))$;\\
        \If{$\supp(P(u_o))\geq\delta$}
        {
            $\kw{keys}[u_o].add(P(u_o))$; /*$P(u_o)$ is a valid \gkey for $u_o$*/ \\
        }
    }
}
\KwRet{\kw{keys}};
\caption{\gkdiscovery($G$, $u_o$, $\mathcal{S}$, $\mathcal{D}$, \kw{keys}, $k$, $\delta$, \kw{size})}
\label{alg:gkdiscovery}
\end{algorithm}
\vspace{-2ex}

The pseudo code of the \gkdiscovery  algorithm is provided in Algorithm~\ref{alg:gkdiscovery}. It is a recursive algorithm to evaluate \emph{k-bounded} \gkeys  for a given type. The algorithm takes as input the graph $G$, a center type $u_o$, summary graph $\mathcal{S}$, (initially empty) dependency graph $\mathcal{D}$, (initially empty) set of \kw{keys}, and three integers $k$, $\delta$, and \kw{size}. The value of \kw{size} is initially set to $0$ and it will be updated for the recursive calls to avoid mining recursive \gkeys  of size greater than $k$. We first create a lattice for the given type $u_o$ (line 1). The function takes $u_o$, the summary graph $\mathcal{S}$ and $\delta$ as an input. It first computes the set of attributes $\mathcal{A}$ and variable nodes $\mathcal{V}$ from $\mathcal{S}$ based on the support parameter $\delta$. It then creates the lattice based on the combination of $\mathcal{A}$ and $\mathcal{V}$. Next, we traverse the lattice in a level-wise manner and check whether each candidate pattern forms a \gkey. Despite traversing the lattice level-wise, the candidate patterns have always height = 1. For each pattern that is not to be pruned (line 3), we first check if it is \emph{k-bounded} (lines 4-5). If the pattern contains a recursive type $t$ without a \gkey, then we need to call the \gkdiscovery  algorithm for $t$. We first check if adding the edge $(u_o, t)$ creates a cycle in $\mathcal{D}$. If so, we remove the edge from $\mathcal{D}$ and remove $t$ from the pattern to avoid cycles in recursive calls (lines 7-10). Otherwise, we call the \gkdiscovery  algorithm by passing $t$ and the current size of the \gkey  (line 12). If we were not able to find a \gkey  for $t$, then we prune the pattern and its descendants (lines 13-14). After these steps, we find the matches of the pattern and compute the number of entities that are uniquely identified by the candidate \gkey  (lines 15-16). We add the pattern as a \gkey  for $u_o$ if it meets the support threshold $\delta$ (lines 17-18). At the end, we return the set of keys that are found.

\subsection{Optimizations} \label{sec:optimization}
In this section, we propose an optimization for the \gkminer  algorithm.
While creating the summary graph $\mathcal{S}$, and as we check the existence of the attributes for each node in $G$, we maintain a hash-map of the values in the attribute domain. This helps us to find which values are unique for each specific attribute. For an attribute $A$, we hash the values $\{a_1,\dots, a_n\}$, where $a_i$ is the value of the attribute $A$ for the node $v_i$ in $G$. The result of the hash is a set of classes $\{\pi_1,\dots, \pi_m\}$, where each $\pi_j$ has one or more equal attribute values, assuming the collision is handled in the hashing process. If a value $a_i$ uniquely exists in a class $\pi_j$, then $a_i$ is a unique value for the attribute $A$ among all the nodes that carry $A$. For each node $v_i \in G$, we may maintain a bit vector flag called $\unique$. We set $v_i.\unique(A)=\true$, if the corresponding value $a_i$ is unique among all nodes that share the same type as $v_i$ and carry attribute $A$. We can use the $\unique$ bit vector when computing the set of matches for $P(u_o)$. Assume $P(u_o)$ contains a set of constant nodes $\{v_{c1}, \dots, v_{cn}\}$. For a match $h \in \mathcal{M}$, if we have $h(u_o).\unique(v_{ci})=\true$ for any attribute $v_{ci}$, then $h$ is uniquely identified by $P(u_o)$ without further exploration. This is true as if an attribute $v_{ci}$ is unique for a node $h(u_o)$ in $G$, then any combination of the attributes that contains $v_{ci}$ is unique for $h(u_o)$. Note that hashing could be done in a constant time. Hence, we maintain the $\unique$ bit vector for all the attributes in $G$ while creating the summary graph $\mathcal{S}$ with the same time complexity $O(V+E)$.

\section{Experiments}
\label{sec:exp}

We use real world graphs to evaluate our algorithm on (1) the efficiency of \gkminer compared to the existing general rule-based mining approach \sakey~\cite{symeonidou2014sakey}; and 
(2) the effectiveness of \gkminer for the task of data linking compared to \sakey.

\stitle{Experimental Setup.} We implement all our algorithms in
Java v17, and ran our experiments on a Linux machine with AMD 2.7 GHz CPU with 128 GB of memory. Our source code and test cases are available online\footnote{\url{https://github.com/mac-dsl/GraphKeyMiner.git}}.

\stitle{Datasets.} We used two real graphs for our experiments.
\begin{enumerate}
    \item \uline{\dbpedia}\cite{lehmann2015dbpedia}:  The graph contains in total 5.04M entities with 421 distinct entity types, and 13.3M edges with 584 distinct labels. \dbpedia is extracted from the Wikipedia pages.
    \item \uline{\imdb}~\cite{imdb}: The data graph contains 6.1M entities with 7 types and 21.3M edges. this dataset contains information of the movies extracted from the IMDB website and in total we have 44.2M facts.
    \newtext{\item \uline{\dbpyago}~\cite{symeonidou2017vickey}: This dataset contains entities from the \dbpedia~\cite{lehmann2015dbpedia} and \yago~\cite{mahdisoltani2014yago3} datasets that are linked together. There exists a gold standard available for the entity links between these two datasets on the \yago Web page~\cite{yagoWebsite}. This dataset uses the ground truth to link the entities across the two knowledge bases. For each entity, we rewrite the properties of the entity in the \yago using its \dbpedia counterparts.}
    
\end{enumerate}

\stitle{Algorithms.} We implemented the following algorithms for the experimental evaluations.
\begin{enumerate}
    \item \uline{\gkminer}: Our mining algorithm of Section~\ref{sec:GKeyDiscovery} with the optimization.
    \item \uline{\gkminerNoOpt}: the \gkminer algorithm without the optimization and usage of the \unique vector.
    \item \uline{\sakey}\cite{symeonidou2014sakey}: Discovers maximal non-keys first and then derive the keys from this set. \sakey does not consider topological constraints to mine keys for graphs.
\end{enumerate}

We excluded \vickey~\cite{symeonidou2017vickey} from our tests as it mines conditional keys over RDFs. \vickey works on top of \sakey by first finding non-keys and then mine conditional keys. As we do not mine conditional graph keys, we do not compare the evaluation of our method with \vickey.

\stitle{Experimental Results.} Firstly, we evaluate the efficiency of \gkminer against \gkminerNoOpt and \sakey. Next, we compare the quality of the mined keys in data linking using the \dbpyago dataset with ground truth~\cite{ma2019ontology}.

\begin{figure}
	\captionsetup[subfloat]{justification=centering}
	\centering
	    {\includegraphics[width=10cm,height=0.47cm]{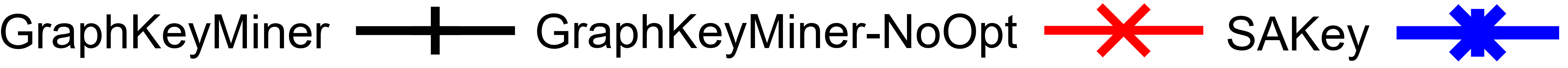}}
		\hfill
		\subfloat[\small{Vary number of types (\dbpedia)}] {\label{fig:gkey_types(dbpedia)}
			{\includegraphics[width=3.8cm,height=2.9cm]{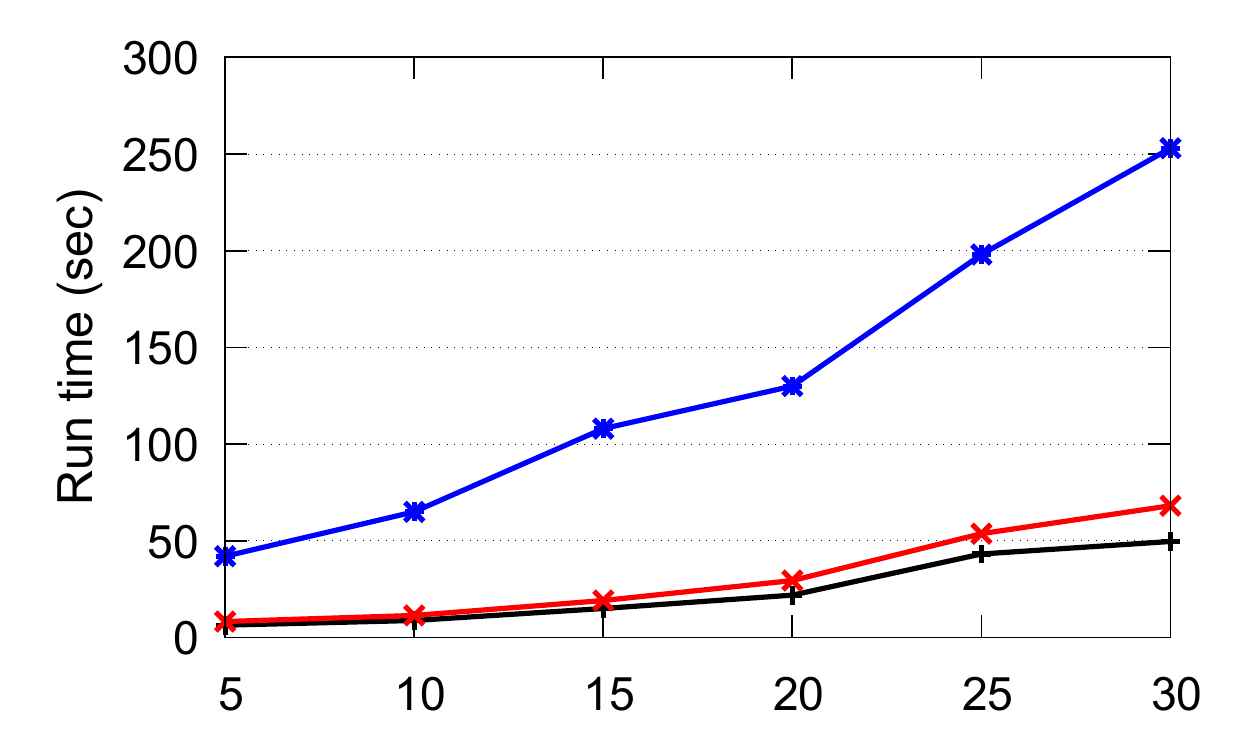}}}
		\hfill\subfloat[
		\small{Vary number of types (\imdb)} ]
		{\label{fig:gkey_types(imdb)}
			{\includegraphics[width=3.8cm,height=2.9cm]{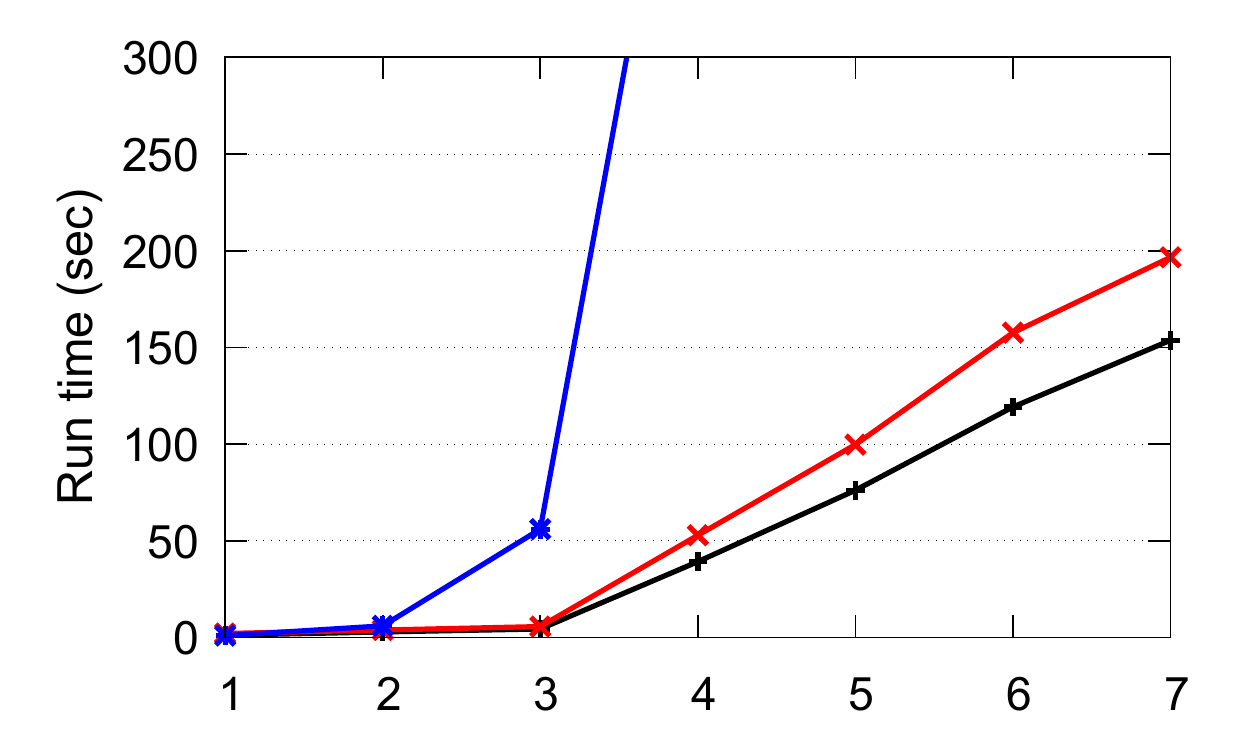}}}
        \hfill\subfloat[
		\small{Vary $k$ (\dbpedia)} 
		]{\label{fig:gkey_k(dbpedia)}
			{\includegraphics[width=3.8cm,height=2.9cm]{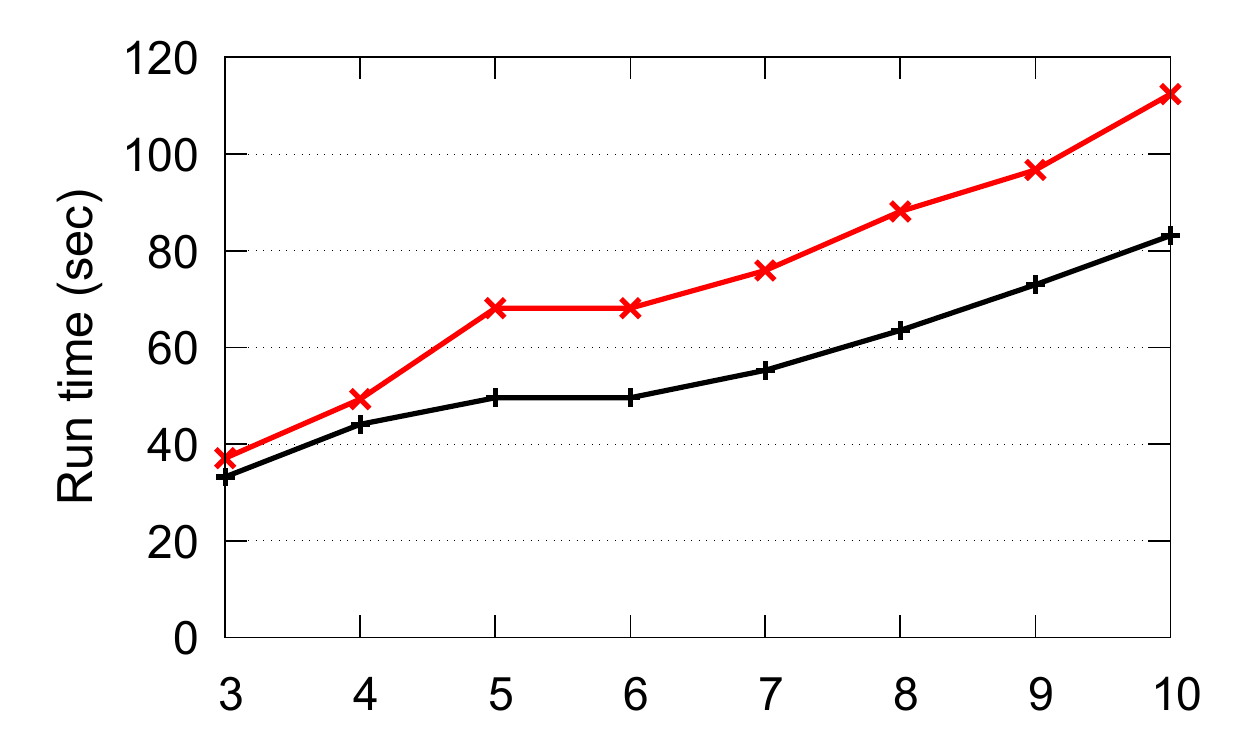}}}
		\hfill\subfloat[
		\small{Vary $k$ (\imdb)}]
		{\label{fig:gkey_k(imdb)}
			{\includegraphics[width=3.8cm,height=2.9cm]{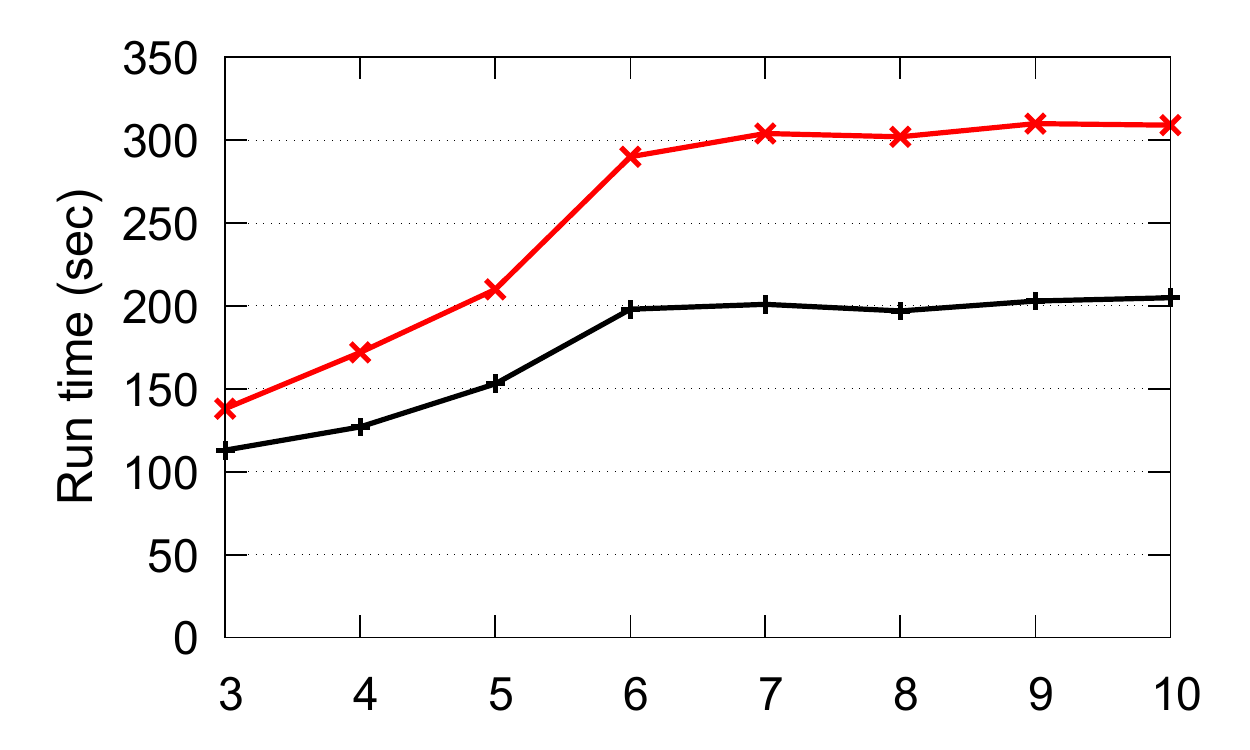}}}
		\hfill \subfloat[\small{Vary \supp (\dbpedia)}]
		{\label{fig:gkey_cov(dbpedia)}
			{\includegraphics[width=3.8cm,height=2.9cm]{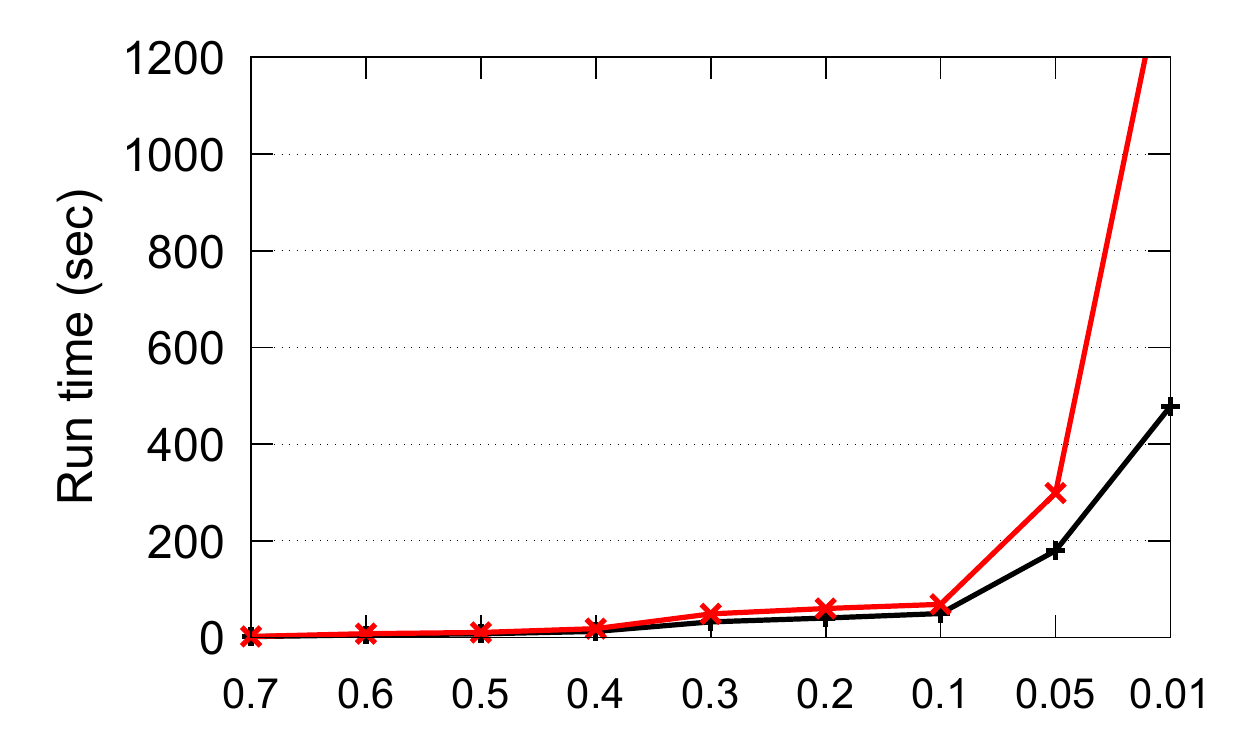}}}		
		\hfill\subfloat[\small{Vary \supp (\imdb)}]
		{\label{fig:gkey_cov(imdb)}
			{\includegraphics[width=3.8cm,height=2.9cm]{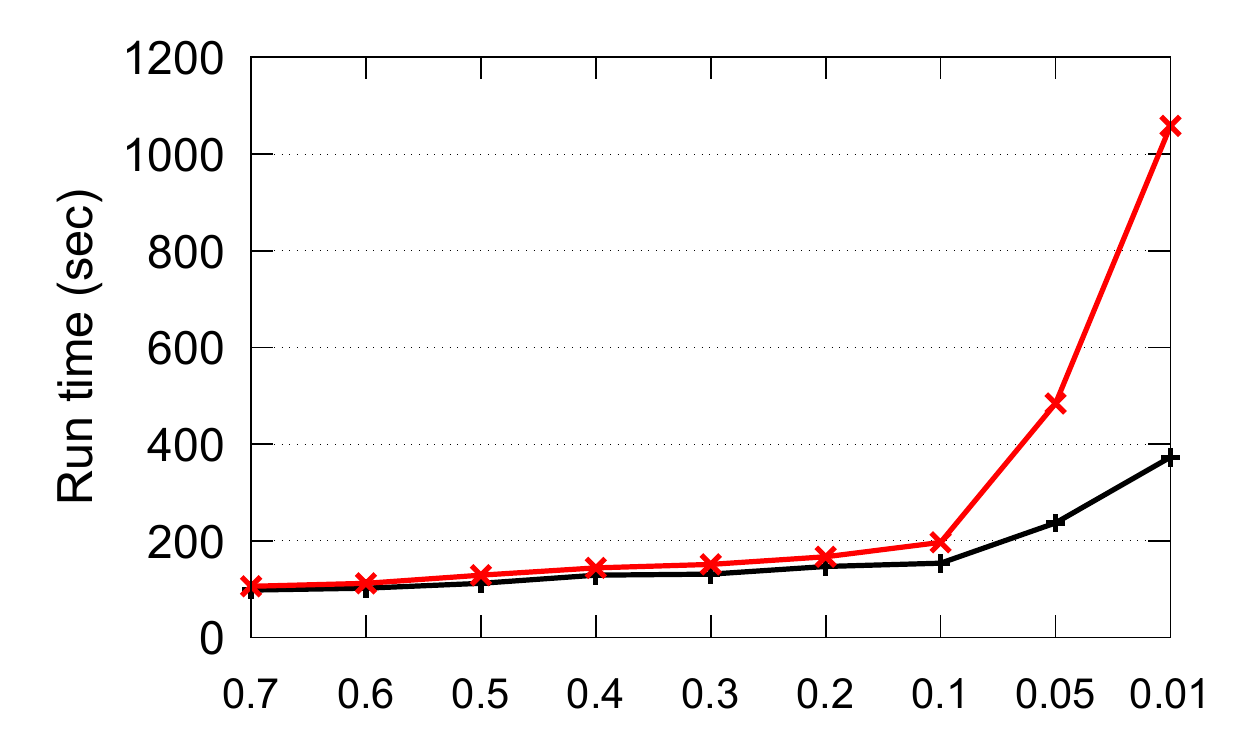}}}
	\caption{\gkminer efficiency.} \label{fig:gkey_efficiency}
\end{figure}

\textbf{Exp-1: Number of types.}All three algorithms take an entity type as input and mine keys for that type. To compare the scalability of the algorithms, we vary the number of types and evaluate the runtime. Using \dbpedia (resp. \imdb) dataset, we fixed the \supp$=10\%$ and $k=5$ and vary the number of types from 5 to 30 (resp. 1 to 7). For \sakey, we set $n=1$ to find exact keys as we do in \gkminer. Figure~\ref{fig:gkey_types(dbpedia)} shows the runtime of the three algorithms. \gkminer is on average 30\% faster than \gkminerNoOpt and 6 times faster than \sakey. This demonstrates the effectiveness of our method with optimizations over the existing method to find graph keys. We stopped executions of \sakey over \imdb after 120 minutes. \sakey was only able to finish mining keys for the types \lit{distributor}, \lit{genre}, and \lit{country} which in total contain only 6.77\% of the facts in the \imdb dataset. However, both \gkminer and \gkminerNoOpt were able to mine \gkeys for all types in less than 200 seconds.

\textbf{Exp-2: Size of pattern.} By fixing \supp$=10\%$, we varied the size of the pattern $k$ from 3 to 10 over \dbpedia and \imdb dataset on 30 and 7 types respectively. We excluded \sakey from this test as there is no pattern size on the keys that \sakey mines. Figure~\ref{fig:gkey_k(dbpedia)} shows the runtime over the \dbpedia dataset, and here is our findings: (1) By increasing the value of $k$, the runtime increases as we have \newtext{larger patterns to match against in $G$}. (2) On average, \gkminer runs 33\% faster than \gkminerNoOpt due to an efficient approach \newtext{find unique values for the attributes that helps to reduce the number of entities to be checked for the  validity of each candidate \gkey}. The same trend exists in the \imdb dataset of Figure~\ref{fig:gkey_k(imdb)}, except the fact that the runtime does not increase for $k>6$. This comes from the fact that we have only 7 types in this dataset and recursion depth (\ie the maximum diameter of the dependency graph) is limited compare to the \dbpedia dataset with over 400 distinct types.

\textbf{Exp-3: Support of \gkey.} In this experiment, we varied the \supp value from $0.01$ to $0.7$ (\ie 1\% to 70\%) on the \dbpedia and \imdb datasets with 30 and 7 types resp., and a fixed pattern size $k=5$. The results are shown in Figure~\ref{fig:gkey_cov(dbpedia)} for \dbpedia and Figure~\ref{fig:gkey_cov(imdb)} for the \imdb dataset. We also excluded \sakey from this test, as there was no option to control the support of a key mined by \sakey. The following is our findings: (1) By increasing the value of \supp, the runtime decreases on both datasets as we have more pruning and fewer number of candidates need to be checked through the lattice. (2) On average, \gkminer runs 66\% and 42\% faster than \gkminerNoOpt on \dbpedia and \imdb respectively. 

\textbf{Exp-4: Effectiveness of \gkminer} To investigate the quality of the \gkeys, we compare the keys mined by \gkminer with the keys of \sakey in the application of entity linking. Primary application of keys is to link entities across two knowledge bases. If two entities are uniquely identified by a key in two different knowledge bases and they share the same attributes, then they refer to the same entity. For this test, we used \dbpyago dataset with the available ground truth~\cite{symeonidou2017vickey}.

\begin{table}[tb!]
\begin{center}
\begin{tabular}{|c|c|c|c|c|c|}
\hline
\multirow{2}{*}{\begin{tabular}{@{}c@{}}\textbf{Entity Type} \\ (\# triples) \end{tabular}}
& \multicolumn{1}{c|}{\gkminer}& \multicolumn{1}{c|}{\sakey}\\ 
\cline{2-3}

\multicolumn{1}{|l|}{\textbf{}}
& \multicolumn{1}{c|}{P/R/F}                                            
& \multicolumn{1}{c|}{P/R/F}
\\ \hline

\multirow{1}{*}{Book(258.4K)}
    
     & 0.99/\textbf{0.07}/\textbf{0.13} & \textbf{1}/0.03/0.06 \\\hline
    
\multirow{1}{*}{Actor(57.2K)}
    
 & \textbf{1}/\textbf{0.36}/\textbf{0.52} & 0.99/0.27/0.43\\\hline
    
\multirow{1}{*}{Museum(12.9K)}
    
     & \textbf{1}/\textbf{0.21}/\textbf{0.34} & \textbf{1}/0.12/0.21 \\\hline
    
\multirow{1}{*}{Scientist(258.5K)}
     & \textbf{0.99}/\textbf{0.09}/\textbf{0.16} & 0.98/0.05/0.11 \\\hline
    
\multirow{1}{*}{University(85.8K)}
    
    & \textbf{0.99}/\textbf{0.12}/\textbf{0.21} & \textbf{0.99}/0.09/0.16 \\\hline
    
\multirow{1}{*}{Movie(832.1K)}
    
    & \textbf{0.99}/\textbf{0.12}/\textbf{0.21} & \textbf{0.99}/0.04/0.08 \\\hline
    
\end{tabular}
\end{center}
\caption{Comparative accuracy of \gkminer against \sakey}
\label{tbl:compare} 
\end{table}

Table~\ref{tbl:compare} shows the \prec (P), \rec(R) and \fscore-score(F) measure of the entity linking task using keys mined by \sakey~\cite{symeonidou2014sakey} against \gkeys mined by \gkminer. Here is our findings: (1) The \prec is always
over 98\% and it is mostly the same in both algorithms. (2) The \rec is low in some cases. This happens as we use a strict string equality when comparing the values of properties. Moreover, the incompleteness of the data in both \yago and \dbpedia leads to lower recall as well. However, the use of recursive keys in \gkminer leads to an increase in recall. For example, for the class \lit{Movie}, \rec increases from 4\% to 12\% when recursive keys are considered. (3) On average, we observe an increase of 7 percentage points in \rec, and of 9 points in \fscore-score using \gkminer against \sakey. This shows the effectiveness of the \gkeys mined by our proposed algorithm \gkminer when we consider recursive keys compared to the classical attribute based keys mined by \sakey.

\vspace{-3ex}
\section{Conclusion and Future Work}
\label{sec:future}
\vspace{-3ex}

We proposed a new algorithm \gkminer to mine graph keys (\gkeys) over real world graphs that is efficient and scalable. We introduce the notion of minimality and support for \gkeys and adapt \gkminer for early termination and pruning of candidate keys.
As next steps, we intend to extend \gkminer to mine conditional \gkeys and study the  the application of conditional \gkeys to data linking, and the parallel discovery of \gkeys in distributed graphs.

\bibliographystyle{splncs04}
\bibliography{main}

\end{document}